\documentclass{article}
\usepackage{graphicx,amssymb}
\begin{document}
\setlength{\textwidth}{27pc}
\setlength{\textheight}{43pc}
\newcommand{\beq}{\begin{equation}}
\newcommand{\eeq}{\end{equation}}
\newcommand{\beqa}{\begin{eqnarray}}
\newcommand{\eeqa}{\end{eqnarray}}
\newcommand{\bs}[2]{{#1}_{\scriptstyle {\rm #2}}}
\newcommand{\bss}[2]{{#1}_{\scriptscriptstyle {\rm #2}}}
\newcommand{\Ho}{\bss{H}{0}}
\newcommand{\Ot}{\bss{\Omega}{T}}
\newcommand{\Oto}{\bss{\Omega}{T,0}}
\newcommand{\Om}{\bss{\Omega}{M}}
\newcommand{\Omo}{\bss{\Omega}{M,0}}
\newcommand{\rc}{\bss{\rho}{crit}}
\newcommand{\rco}{\bss{\rho}{crit,0}}
\newcommand{\rl}{\bss{\rho}{\Lambda}}
\newcommand{\Ol}{\bss{\Omega}{\Lambda}}
\newcommand{\Olo}{\bss{\Omega}{\Lambda,0}}
\newcommand{\Rl}{\bss{R}{\Lambda}}
\newcommand{\rt}{\bss{\rho}{T}}
\newcommand{\Otd}{\bss{\Omega}{T,d}}
\newcommand{\Otn}{\bss{\Omega}{T,n}}
\newcommand{\ao}{\bss{a}{0}}
\newcommand{\ad}{\bss{a}{d}}
\newcommand{\an}{\bss{a}{n}}
\newcommand{\zd}{\bss{z}{d}}
\newcommand{\td}{\bss{t}{d}}
\newcommand{\tn}{\bss{t}{n}}
\newcommand{\xo}{\bss{x}{0}}
\newcommand{\adotao}{\bss{(\dot{a}/a)}{0}}
\newcommand{\uH}{\bss{u}{H}}
\newcommand{\fmin}{\bss{f}{min}}
\newcommand{\uHo}{\bss{u}{H,0}}
\newcommand{\Vmin}{\bss{V}{min}}

{\LARGE {\bf \noindent The Nearly Flat Universe}}

\vspace{15mm}

{\large
{\bf \noindent R.J. Adler\footnote[1]{Gravity Probe B, Hansen Experimental Physics Laboratory, Stanford University, Stanford CA 94305-4085 U.S.A.}$^{,}$\footnote[2]{Corresponding author.  E-mail: adler@relgyro.stanford.edu}}
and
{\bf J.M. Overduin$^{1,}$\footnote[3]{E-mail: overduin@relgyro.stanford.edu}}
}

\begin{abstract}
We study here what it means for the Universe to be nearly flat, as opposed to
exactly flat.  We give three definitions of nearly flat, based on density,
geometry and dynamics; all three definitions are equivalent and depend on
a single constant flatness parameter $\varepsilon$ that quantifies the notion
of nearly flat.  Observations can only place an upper limit on $\varepsilon$,
and always allow the possibility that the Universe is infinite with $k=-1$ or
finite with $k=1$.  We use current observational data to obtain a numerical
upper limit on the flatness parameter and discuss its implications, in
particular the ``naturalness'' of the nearly flat Universe.
\end{abstract}

\begin{center}
\end{center}

\section{Introduction} \label{sec:int}

\indent Observational cosmologists tell us that, at present, the ratio of
the sum of the densities of all forms of matter-energy in the Universe to
the critical density required for spatial flatness is:
\beq
\Oto = 1.02 \pm 0.04 \; .
\label{WMAP}
\eeq
We follow standard usage throughout, using the symbol $\Omega$ to denote
densities relative to the critical density $\rc\equiv(3/8\pi G)(\dot{a}/a)^2$,
with $\rl=\Lambda/8\pi G$ representing the contribution from vacuum
energy so that $\Ot=(\rho+\rl)/\rc$.  The subscript ``0'' indicates terms
taken at the present time.

Eq.~(\ref{WMAP}) is a 95\% confidence-level fit to combined data on
fluctuations in the spectrum of cosmic microwave background (CMB) radiation
from the WMAP satellite and observations of the magnitudes of distant
Type~Ia supernovae (SNIa), plus the assumption that the present value
of Hubble's parameter satisfies 
$\Ho=\adotao>50$~km~s$^{-1}$~Mpc$^{-1}$ \cite{Spe03}.
The data are consistent with the possibility that $\Oto$ is {\em precisely\/}
one, implying an infinite and spatially flat ($k=0$) Universe.
They are also marginally consistent with the idea, widely held during
the 1990s, that $\Oto<1$, implying a Universe which is infinite and
negatively curved ($k=-1$).  Most intriguingly, 
Eq.~(\ref{WMAP}) suggests the possibility of a closed, finite and positively
curved Universe with $\Oto>1$ and $k=+1$ --- a prospect which has received
less theoretical attention over the years \cite{Fel86,Whi96,Kam96,Ove01}
but which is now hinted at by both the location of the primary acoustic
peak \cite{Whi00,del03} and the amplitude of low-order multipoles
\cite{Efs03,Uza03} in the CMB power spectrum.  These are three radically
different models of the Universe (Fig.~1).
\begin{figure}[t!]
\begin{center}
\includegraphics[width=\textwidth]{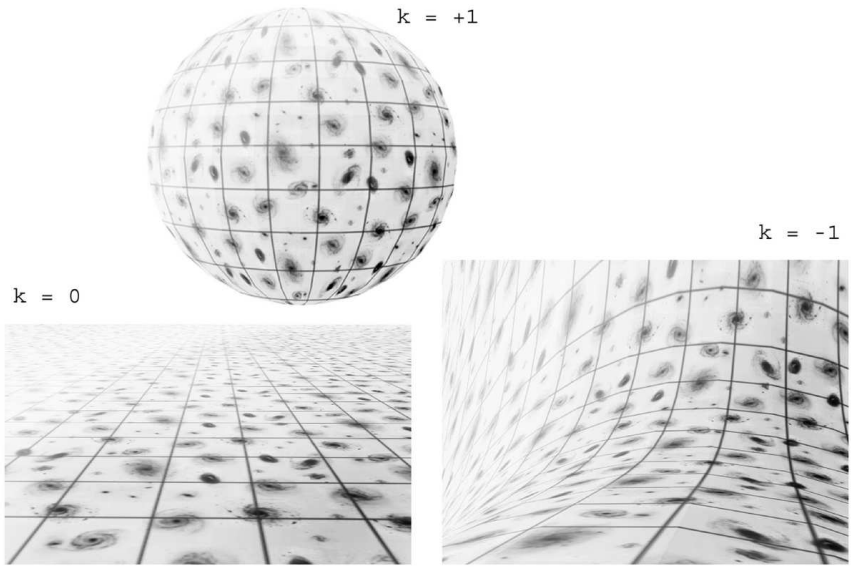}
\end{center}
\caption{Closed ($k=+1$), flat ($k=0$) and open ($k=-1$) universes.}
\end{figure}

The flat $k=0$ Universe is widely favored among cosmologists for two reasons.
First, inflation dynamically causes a very small portion of the primordial
Universe to grow so large as to encompass the entire observable Universe, 
so that the spatial structure of the Universe is greatly flattened.  This,
however, only forces the Universe to be nearly, rather than exactly flat
\cite{KT90,Pee93}, and indeed there are examples of inflationary models
with $k\neq0$ \cite{Ell88,Hub91,Lin03,Pav03,Ell04,Las04}.  Secondly, it is
felt by some that a nearly flat Universe involves a numerical coincidence
or fine-tuning between $\Ot$ and unity, especially at early times.  As we will
discuss in \S6, however, this apparent fine-tuning is a natural result of the
definitions of the density ratio $\Ot$ and the critical density, combined
with the cosmological equations.  We must also mention that notions of
naturalness and fine-tuning are subjective; for example, one might consider
a flat Universe to be infinitely fine-tuned since it has $\Ot$ identically
equal to one, thereby making it the most unnatural choice.

Our main purpose in this work is to define just what it means for the
Universe to be nearly flat.  We give three logically independent but
physically equivalent definitions, based on density, geometry and dynamical
behavior.  All lead to the same dimensionless constant, whose significance
as a flatness parameter does not appear to have been widely appreciated.
Explicitly, this parameter is the ratio of the de~Sitter radius of the
Universe to a constant of integration in the Friedmann-Lema\^{\i}tre equation.

We derive our flatness parameter on the basis of density in \S3,
geometry in \S4, and dynamics in \S5.  The question of naturalness 
is addressed in \S6, and in \S7 we give our summary and conclusions.

Our discussion has features in common with that of Chernin \cite{Che03}.
Chernin discusses four parameters, which he calls Friedmann constants; these
are $A_V$ associated with vacuum energy, $A_D$ associated with dark matter,
$A_B$ associated with baryons, and $A_R$ associated with radiation.  He
emphasizes that all are very roughly equal (within a few orders of magnitude),
and suggests that this may be evidence of a cosmic symmetry of unknown origin.
On the other hand, we emphasize that for a nearly flat Universe, $A_V$ must
be at least several orders of magnitude less than $A_D$, and that indeed this
is responsible for the near flatness.  Thus our work is consistent with and
complementary to that of Chernin.

The issue of near-flatness is also raised in a recent article by Lake
\cite{Lak04}, who adopts a rather different approach from ours, based on
the trajectories of model universes in the $\Om-\Ol$ plane.  Lake's intent
is to assess the likelihood that the Universe is exactly flat, whereas we
would contend that observation cannot distinguish --- even in principle ---
between a perfectly flat Universe and one that is sufficiently close to flat.

\section{Cosmological Dynamics}

We begin by briefly reviewing cosmological dynamics, focusing on the present
epoch in which the Universe is dominated by a cosmological constant (or dark
energy) and pressureless matter.  The matter component could be further broken
down into cold dark matter and baryonic matter, but these have the same effect
on the dynamics and we treat them together.  For early times, radiation is
also important \cite{Che03}.  We assume a linear equation of state for the
matter component, $p=w\rho$ with $w=$~const.  The basic equations of
homogeneous and isotropic cosmology may be written (with $c=1$):
\beq
8\pi G\rho = -\Lambda + 3 \left( \frac{k}{a^2}+\frac{\dot{a}^2}{a^2} \right) ,
   \hspace{1cm}
8\pi G p = \Lambda - \left( \frac{k}{a^2}+\frac{\dot{a}^2}{a^2} +
   \frac{2\ddot{a}}{a} \right) \; .
\label{Ron1}
\eeq
With the linear equation of state and some algebraic manipulation, we obtain
a first integral of these (the Friedmann-Lema\^{\i}tre equation) as
\beq
\dot{a}^2 = \left( \frac{C}{a} \right)^{3w+1} + \frac{a^2}{\Rl^2} - k \; ,
\label{Ron2}
\eeq
where $C$ is a constant and $\Rl$ is the de~Sitter radius, defined by:
\beq
\Rl^2 = \frac{3}{\Lambda} 
    = \frac{3}{8\pi G \rl} 
    = \frac{1}{H^2 \Ol} 
    = \frac{1}{\Ho^2 \Olo} \; .
\label{RlDefn}
\eeq
Here $\Olo=\rl/\rco$ is the present density ratio of vacuum energy; we note
that the vacuum energy density is constant but the density ratio $\Ol$ is not.
(The constant $C$ here is the same as in Lake's paper \cite{Lak04},
while $C$ and $\Rl$ are referred to as $A_D$ and $A_V$ respectively in
the paper by Chernin~\cite{Che03}.)  Most of the discussion in this paper
will be based on Eq.~(\ref{Ron2}).  For the case of a flat ($k=0$) Universe,
it can be solved exactly when the matter component is dominated by, e.g.,
pressureless dust ($w=0$) or radiation ($w=1/3$):
\beq
a(t) = \left\{ \begin{array}{cc}
   (C\Rl^2)^{1/3} \sinh^{2/3} (3t/2\Rl) & (w=0) \\
   (C\Rl)^{1/3} \sinh^{1/2} (2t/\Rl) & (w=1/3)
\end{array} \right. \; .
\label{CM}
\eeq
The first of these expressions governs the now-standard concordance model
of present-era cosmology (see Fig.~2).  It is found in surprisingly few of
the standard texts \cite{Rin77,Pad93} and goes back to Heckmann in 1931
\cite{BHP97,Hec31}.
\begin{figure}[t!]
\begin{center}
\includegraphics[width=6.6cm]{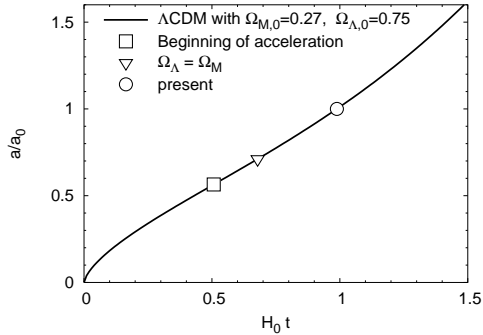}
\end{center}
\caption{Scale factor as a function of time in the standard concordance model.}
\end{figure}

Some special times are of particular interest in this model:
the time when acceleration begins, the time of equality between energy
densities of matter and dark energy, and the present time (when dark-energy
density is approximately 2.8 times the matter density).  These are conveniently
characterized by:
\beq
\sinh(3t/2\Rl) = \left\{ \begin{array}{cl}
   1/\sqrt{2} & \mbox{ (acceleration begins)} \\
   1 & \mbox{ (equal matter/dark-energy density)} \\
   \sqrt{\Olo/\Omo} & \mbox{ (at present, $\approx 1.67$)}
\end{array} \right. \; .
\label{Ron4}
\eeq
Best-fit values of the relevant cosmological parameters from the WMAP
satellite are
$\Ho=71$~km~s$^{-1}$~Mpc$^{-1}, \Omo=0.27$ and $\Olo=0.75$ \cite{Spe03}, so
that the de~Sitter radius works out to $\Rl=16$~Gly from Eq.~(\ref{RlDefn}).

Note that for the case $k=0$
the equations are scale-invariant, so the scale function is arbitrary
to within a multiplicative factor and is not a measurable quantity.
For $k=\pm1$ the scale function is measurable in terms of $H$ and $\Ot$,
as we will show in the next section.

\section{Density Definition of Nearly Flat}

Since density can never be measured with perfect precision, it is clear 
that we cannot use the quantity $\Oto$ itself to verify whether or not
the Universe is precisely flat.  But the observational bound~(\ref{WMAP})
certainly indicates that the Universe is either flat, or {\em nearly\/} so.
In fact, we will use $\Ot$ to obtain our first definition of nearly flat.

The first of the cosmological field equations~(\ref{Ron1}) together with
the first integral~(\ref{Ron2}) allows us to solve for $\Ot$ in terms of
the scale factor as follows:
\beq
\Ot = \left[ 1 - \frac{k}{(C/a)^{3w+1} + a^2/\Rl^2} \right]^{-1} =
      \left[ 1 - \frac{k}{f(a)} \right]^{-1} \; ,
\label{Ron6}
\eeq
where the function $f(a)$ reads
\beq
f(a) \equiv (C/a)^{3w+1} + a^2/\Rl^2 = C/a + a^2/\Rl^2
\eeq
for the present era with $w=0$.  We plot this function in Fig.~3
\begin{figure}[t!]
\begin{center}
\includegraphics[width=6.6cm]{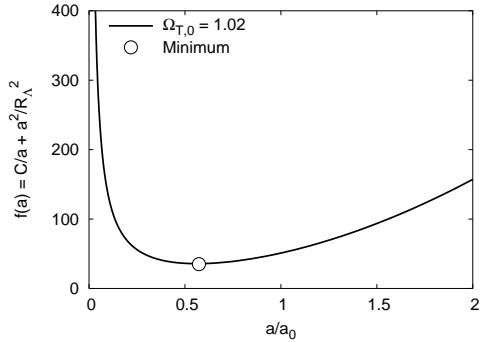}
\end{center}
\caption{Behavior of the function $f(a)$ that determines the density ratio,
   with minimum $\fmin=3(C/2\Rl)^{2/3}$ at $a=(C\Rl^2/2)^{1/3}$.  We have
   used Eqs.~(\ref{RlDefn}) and (\ref{Cdefn}) for $\Rl$ and $C$ respectively,
   with WMAP values for $\Omo$ and $\Omo$.}
\end{figure}
and note that its minimum value is $3(C/2\Rl)^{2/3}$.
The total density parameter itself is plotted in
Fig.~4; note that $\Ot\rightarrow1$ in the early Universe when
$a\rightarrow0$, and that $\Ot\rightarrow1$ in the distant future when
$a$ becomes very large.
Since observations indicate that $\Ot$ is close to one, the second term
in square brackets must be small, so we may expand as
\beq
\Ot \approx 1+\frac{k}{C/a+a^2/\Rl^2} \; .
\label{Ron7}
\eeq
The parameter $\Ot$ will be close to unity if
\beq
f(a) = C/a+a^2/\Rl^2 \; > \; \fmin = 3(C/2\Rl)^{2/3} \; \gg \; 1 \; .
\label{Ron8}
\eeq
A nearly flat Universe with a total density parameter $\Ot$ that is nearly
equal to unity for all times is therefore equivalent to one with a small
{\em flatness parameter} $\varepsilon$, defined as
\beq
\varepsilon \equiv \Rl/C \ll 1 .
\label{epsilonDefn}
\eeq
This parameter is the constant ratio of a fundamental constant in the
theory $(\Rl$) to a constant of integration ($C$), so it is allowed by the
theory to have any value.  (It is equal to Chernin's $A_V/A_D$ \cite{Che03}.)
However, many theorists would consider it unnatural for $\varepsilon$ to
differ from unity by many orders of magnitude, or for $C$ to differ by
many orders of magnitude from the natural value $\Rl$.

\begin{figure}[t!]
\begin{center}
\includegraphics[width=6.6cm]{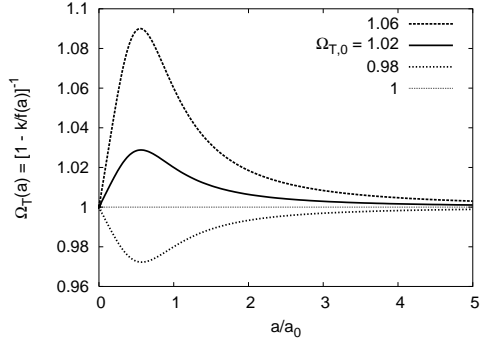}
\end{center}
\caption{Behavior of the density ratio versus scale factor, using
   Eqs.~(\ref{RlDefn}) and (\ref{Cdefn}) for $\Rl$ and $C$ respectively,
   and assuming WMAP density parameters at present ($a/\ao=1$):
   $\Oto=1.02\pm0.04$.
   The flat case $\Oto=1$ is also shown for comparison.}
\end{figure}

The constant $C$ is simply related to the matter density; this can be seen
by comparing the first of Eqs.~(\ref{Ron1}) with Eq.~(\ref{Ron2}) divided
by $a^2$ (for $w=0$):
\beq
C=(8\pi G/3)(\rho a^3) = \Om H^2 a^3 = \Omo \Ho^2 \ao^3 ,
\label{Cdefn}
\eeq
where the final steps follow from the definition of $\Om$ and the fact
that $C=$~constant.  This relation expresses the well-known way in which
matter density decreases as the Universe expands.

Let us assume that the Universe is not exactly flat, but is nearly flat,
and use the observational bound~(\ref{WMAP}) to obtain an upper limit and
rough estimate for the flatness parameter.  If we divide the first of
Eqs.~(\ref{Ron1}) by $3H^2$ we obtain
\beq
\Ot-1 = k/H^2 a^2 \; .
\label{Ron12a}
\eeq
Combining our expressions~(\ref{RlDefn}) and (\ref{Cdefn}) with this,
we find that
\beq
\varepsilon = \left( \Ho^3 \ao^3 \Olo \sqrt{\Oto} \right)^{-1}
= \frac{|\Oto-1|^{3/2}}{\Omo \sqrt{\Olo}} \lesssim 0.012 \; .
\label{Ron12b}
\eeq
It then follows from the definitions~(\ref{RlDefn}) and (\ref{epsilonDefn})
that the constant $C$ must be at least
\beq
C = \frac{1}{\varepsilon \Ho \sqrt{\Olo}} \gtrsim \mbox{ 1300 Gly } \; .
\label{Ron13a}
\eeq
Moreover the scale function can be determined explicitly if $k\neq0$;
from Eq.~(\ref{Ron12a}) it is given by
\beq
a = \left( \frac{k}{\Ot-1} \right)^{1/2} \frac{1}{H} \; ,
\label{Ron13b}
\eeq
which takes the value $\ao\approx99$~Gly at present.  (This result nicely
dislays how $a$ becomes indeterminate for $k=0$ and $\Ot=1$.)  The value of
the flatness parameter in Eq.~(\ref{Ron12b}) is small, but can hardly be
considered unnatural.  Similarly comments apply to the distances in
Eq.~(\ref{Ron13a}) and (\ref{Ron13b}).

It is amusing to note that an independent means of determining $\varepsilon$ 
is available in principle.  From the first of Eqs.~(\ref{Ron2}) we may obtain
\beq
(H/\Ho)^2 = \Omo(\ao/a)^3 + \Olo - (\Oto-1)(\ao/a)^2 \; .
\eeq
For $k=1$, $(H/\Ho)^2$ dips below its asymptotic value of $\Olo$, dropping
to a minimum of $(H_{\ast}/\Ho)^2=[1-(4/27)\varepsilon^2]\Olo$ at
$(a_{\ast}/\ao)=3\Omo/2(\Oto-1)$.  Thus observations of the expansion
rate at this epoch would yield a definitive measurement of  $\varepsilon$.
Unfortunately the relevant time occurs so far in the future
(of order 60~Gyr) that such a method is not of practical interest.

In summary, we may say that, in terms of density, the Universe is nearly
flat if the flatness parameter defined in Eq.~(\ref{epsilonDefn}) is small,
which means that the constant of integration $C$ is large in comparison to
the de~Sitter radius $\Rl$.  Future improvements in measurements of the density
ratio $\Oto$ are obviously of great interest, particularly since its current
value hints at a density higher than the critical one, implying that the
Universe may have positive spatial curvature and be finite.

\section{Geometrical Definition of Nearly Flat}

Geometry provides the most intuitive method for defining near-flatness.
A part of the surface of a sphere is nearly flat if its extent is small
compared to the radius of the sphere.  In the same way, we may say that
the Universe is nearly flat if the region to which we have access is
small compared to its characteristic radius.  For a metric in the
standard form,
\beq
ds^2 = dt^2 - a^2(t) \left( \frac{du^2}{1-k u^2} + u^2 d\theta^2 +
   u^2 \sin^2\!\theta \, d\phi^2 \right) \; ,
\eeq
this may be expressed mathematically by saying that the Universe is flat
within some region (i.e., within proper radius $\ell$) if its dimensionless
comoving radial coordinate $u$ remains small throughout that region:
\beq
u^2 \ll 1 \; .
\label{Ron14}
\eeq
Our region of interest is the accessible Universe, so we take $\ell$ to be
the Hubble distance $\ell=c/H$.

Proper (or physical) distance $\ell$ is calculated in terms of $u$ by
\beqa
\ell = \int_0^u \frac{a(t)\,du^{\prime}}{\sqrt{1-k{u^{\prime}}^2}}
   & = & \left\{ \begin{array}{ll}
     a(t) \sin^{-1}\!u & (k=1) \\
     a(t) u & (k=0) \\
     a(t) \sinh^{-1}\!\!u & (k=-1)
\end{array} \right. \nonumber \\
   & = & a(t) u [ 1 + \textstyle{\frac{k}{6}}u^2 + \cdots \; ] \; .
\label{ellDefn}
\eeqa
Eq.~(\ref{ellDefn}) is easily inverted for $u=u(\ell)$:
\beq
u = (\ell/a) [ 1 - \textstyle{\frac{k}{6}}(\ell/a)^2 + \cdots \; ]
   \; .
\eeq
We will need only the lowest order in $(\ell/a)$, so that the value of $k$
is irrelevant.  Taking $\ell=1/H=a/\dot{a}$ for the accessible Universe,
we find for the square of the comoving Hubble distance:
\beq
\uH^2 = (1/\dot{a})^2 = (C/a + a^2/\Rl^2 - k)^{-1} 
      = [f(a)-k]^{-1} \; .
\label{uHdefn}
\eeq
The equivalence of the definitions in terms of density and geometry becomes
clear if we use Eqs.~(\ref{Ron6}) and (\ref{uHdefn}) to relate $\Ot$ to $\uH$,
as follows:
\beq
\uH^2 = \frac{\Ot-1}{k} \; .
\eeq
That is, the comoving Hubble distance and the deviation of the density ratio
from unity are small together, and both vanish at very early and very late
times.  Using the WMAP value $\Oto\approx1.02$, the present spatial extent
of the Universe in terms of the comoving Hubble distance is 
$\uHo\approx\sqrt{0.02}=0.14$.

In brief summary, the geometric definition of near-flatness is completely
equivalent to the density definition.

\section{Dynamical Definition of Nearly Flat}

The first integral~(\ref{Ron2}) may be put into dimensionless form by the
substitutions $\tau\equiv t/\Rl$ and $x\equiv a/(C\Rl^2)^{1/3}$ to obtain
\beq
(dx/d\tau)^2 - (1/x + x^2) = -k\varepsilon^{2/3} \; .
\label{Ron19}
\eeq
It is apparent that, for small $\varepsilon$, the dynamical behavior will be
similar to that for $k=0$.  Thus, as in the previous sections, we are led to
$\varepsilon$ as a measure of near-flatness.  Eq.~(\ref{Ron19}) can of course
be solved exactly by quadratures, as we will do below, but for a perturbative
solution the appropriate expansion parameter is clearly $k\varepsilon^{2/3}$.

\begin{figure}[t!]
\begin{center}
\includegraphics[width=6.6cm]{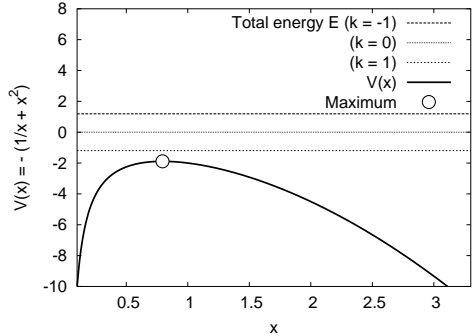}
\end{center}
\caption{The mechanical analogy for the dynamics of the Universe for each
   of the three cases of spatial curvature.  The maximum of the potential
   occurs at $x=1/2^{1/3}$ where $\Vmin=-3/2^{2/3}$.}
\label{MechAnalogy}
\end{figure}

It is worth noting the mechanical analogy for Eq.~(\ref{Ron19}), which is 
a particle of mass 2 and total energy $E=-k\varepsilon^{2/3}$ in a potential
$V(x)=-(1/x+x^2)$, as shown in Fig.~\ref{MechAnalogy}.
The particle flies rapidly outward from the origin, coasts over the top of 
the potential hill, and accelerates down the hill to infinity.  Obviously,
if the total energy $E$ is much smaller than the distance between the top of
the hill to zero (which is $3/2^{2/3}$), then total energy is unimportant
and the three cases $k=0,\pm1$ will exhibit the same qualitative behavior.
The difference is a shift in the time required to reach a given value of $x$.

Let us illustrate the use of Eq.~(\ref{Ron19}).  Very early in the history 
of the Universe, its expansion is dominated by the $1/x$ term, while for its
very late history it is dominated by the $x^2$ term.  Only at intermediate
times is the last perturbative term relevant.  This is easily seen if we
take the positive root of Eq.~(\ref{Ron19}) and solve exactly by quadratures,
then expand in $\delta\equiv k\varepsilon^{2/3}$ as follows:
\beq
\tau = \int_0^x \frac{dx}{\sqrt{1/x + x^2 - \delta}} =
   \sum_{n=0}^{\infty} \delta^n c_n I_n \; .
\label{Quad}
\eeq
The integrals $I_n$ are given by
\begin{equation}
I_n \equiv \int_0^x \frac{dy}{(1/y + y^2)^{n+1/2}} \; ,
\label{InDefn}
\end{equation}
with $c_0\equiv1$ for $n=0$ and $c_n\equiv\frac{1\cdot3\cdots(2n-1)}
{2\cdot4\cdots(2n)}$ for $n>0$.

To zeroth order in $\delta$ one has
\beq
\tau = I_0 = (2/3)\ln(x^{3/2} + \sqrt{1+x^3}) \; .
\eeq
This is equivalent to the first of Eqs.~(\ref{CM}), valid for $k=0$.
The higher-order terms in the expansion~(\ref{Quad}) represent the shift in
time required for the Universe to attain a given value of $a$ when $k=\pm1$,
relative to the $k=0$ case.  The integrals $I_n$ cannot be expressed in terms
of elementary functions for $n>0$.  However, a change of variables to
$s\equiv y^3$ allows one to rewrite them in terms of the hypergeometric
function $F(a,b;c,z)$, following Silbergleit \cite{Sil02}:
\beqa
I_n & = & \textstyle{\frac{1}{3}} \displaystyle\int_0^{x^3} (1+s)^{-(n+1/2)} 
   s^{n/3-1/2} ds \nonumber \\
    & = & \textstyle{\frac{\displaystyle{x^{n+3/2}}}{n+3/2}}
   F(n+\textstyle{\frac{1}{2}},
   \textstyle{\frac{n}{3}}+\textstyle{\frac{1}{2}};
   \textstyle{\frac{n}{3}}+\textstyle{\frac{3}{2}},-x^3) \; .
\eeqa
In particular, it is possible to solve exactly for the accumulated time shift
very late in the history of the Universe using Eq.~(\ref{InDefn}).
In the limit as $a$ goes to infinity, one has:
\beq
\lim_{x\rightarrow\infty}I_n = \int_0^{\infty} 
   \frac{dy}{(1/y + y^2)^{n+1/2}} = \frac{
   \Gamma(\textstyle{\frac{n}{3}}+\textstyle{\frac{1}{2}})
   \Gamma(\textstyle{\frac{2n}{3}})}{
   3\Gamma(n+\textstyle{\frac{1}{2}})} \; .
\eeq
The leading contributions to $\Delta\tau\equiv\tau-I_0$ are then found 
(in the late-time limit) as
\beqa
\lim_{a\rightarrow\infty}\Delta\tau & = &
   \textstyle{\frac{1}{2}} (k\varepsilon^{2/3})
   \displaystyle{\lim_{x\rightarrow\infty}I_1} +
   \textstyle{\frac{3}{8}} (k\varepsilon^{2/3})^2 
   \displaystyle{\lim_{x\rightarrow\infty}I_2} + \cdots \nonumber \\
& = & 0.29 \,k \varepsilon^{2/3} + 0.075 \,k^2 \varepsilon^{4/3} \cdots
\eeqa
In conjunction with the upper limit~(\ref{Ron12b}) on $\varepsilon$,
the contributions of these two terms to the asymptotic shift in physical
time $\Delta t=\Rl\Delta\tau$ amount to no more than $\pm240$~Myr and
3~Myr respectively.

The time shift $\Delta\tau$ that has accumulated up to the present time
can be calculated by evaluating Eq.~(\ref{InDefn}) numerically with 
$x=\xo=(\Olo/\Omo)^{1/3}$; one finds in this way that the first two 
terms in the expansion contribute no more than $\pm120$~Myr and
2~Myr, respectively.  These times are small on cosmological scales,
but not negligible.  For instance, they are much longer than the duration
of the radiation-dominated era.

\section{Flatness of the Early Universe}

At present the Universe is either flat with $k=0$, or nearly flat with
flatness parameter $\varepsilon\leqslant10^{-2}$.  But the early Universe
was extremely close to density-flat, as emphasized by many authors,
notably Dicke \cite{Dic70}.  Some authors have suggested that $k=0$ is
the most natural choice, since otherwise the density ratio $\Ot$ must have
been fine-tuned to be very close to unity, to within about a part in $10^{17}$
at the time of nucleosynthesis.  This is the ``flatness problem'' of
standard cosmology, and one of the issues that the inflationary paradigm
is claimed to address \cite{KT90}.

Here we advocate a rather different view, for two reasons.  First, as 
mentioned in the introduction, the notion of naturalness is fundamentally
subjective, and one is equally entitled to view the case $k=0$ as
{\em infinitely\/} fine-tuned, with $\Ot$ identically equal to one.
Second, as we will show below, the difference between the actual density
and the critical density actually diverges like $t^{-1}$ in the early
radiation-dominated Universe.  The density ratio $\Ot$ approaches
unity only because the critical density diverges even faster, like $t^{-2}$.
The apparent fine-tuning is thus inherent in the definition of the critical
density and the equations of general relativity, in particular
Eq.~(\ref{Ron6}).  As we will discuss in detail below, $\Ot$ {\em must\/}
approach very close to one at early times.  Thus the quantity $\Ot-1$ is
not a useful measure of naturalness or fine-tuning, and we suggest the
use of the constant flatness parameter $\varepsilon$ instead.  Since this
is as large as $10^{-2}$ according to present observations, a nearly
flat Universe cannot as yet be claimed to be excessively fine-tuned.

Let us first calculate the deviation of the total density from critical in
the early Universe.  From the first of the cosmological equations~(\ref{Ron1})
and the definition of the critical density, we see that
\beq
\rt-\rc=(3/8\pi G) (k/a^2) \; .
\label{Ron31}
\eeq
In the early radiation era, $a\propto t^{1/2}$ so for $k\neq0$ the difference
in Eq.~(\ref{Ron31}) diverges at early times like $t^{-1}$.  The quantity
$\Ot-1=(\rt-\rc)/\rc$ is small only because the critical density diverges
even more rapidly:
\beq
\rc=(3/8\pi G) (\dot{a}/a)^2 \propto t^{-2} \; .
\eeq
The smallness of the quantity $\Ot-1$ at early times is thus an artifact
of the definition of the critical density.

Let us next estimate the deviation of the density ratio from unity at the
time of decoupling, which occurred at a redshift of about 
$\zd\approx(\ao/\ad)\approx10^3$ at a time $t\approx10^{12}$~s; this is
also roughly the time of matter and radiation density equality.  From
Eq.~(\ref{Ron6}):
\beq
\Ot-1 \approx k/f(a) \approx ka/C \; .
\eeq
This is a reasonable approximation in the matter-dominated era (i.e., until
recently), during which $a\propto t^{2/3}$.  Thus the deviation of $\Ot$ from
unity at decoupling obeys
\beq
(\Otd-1)/(\Oto-1) \approx \ad/\ao \approx 1/\zd \approx 10^{-3} \; .
\eeq
Since the present deviation is about $\leqslant 10^{-2}$ we obtain a
small deviation of about $\lesssim 10^{-5}$ at decoupling.

Similarly, we may go back in time in the radiation era, during which
$w=1/3$ and $a\propto t^{1/2}$, to a time of about $\tn\approx 1$~s,
when nucleosynthesis occurred.  As above, but with $w=1/3$ in 
Eq.~(\ref{Ron6}), we obtain
\beq
(\Otn-1)/(\Otd-1) \approx (\an/\ad)^2 \approx \tn/\td \approx 10^{-12} \; .
\eeq
Thus the deviation during nucleosynthesis was only about $10^{-17}$, an
impressively small number.  It follows merely from the present density
deviation, which is not necessarily very small, plus the cosmological
equations and the definition of the critical density.

A useful analogy from elementary physics might be the following: consider a
test particle of mass $m$ with total energy $E$ falling into the Newtonian
gravitational field of a mass $M$.  The ratio of this particle's kinetic
energy $K=mv^2/2$ to its potential energy $|U|=GMm/r$ is $K/|U|=(E/GMm)r+1$.
Note that the difference $K/|U|-1$ becomes arbitrarily small as one
approaches $r\rightarrow0$, in exactly the same way that $\Ot-1$ does in
cosmology as $t\rightarrow0$.  Yet one would hardly be justified in
concluding from this that $E$ ``must be'' zero on the grounds of naturalness.

In summary, the extremely small deviation of the density ratio from unity
in the early Universe is a consequence of the definition of the critical
density and the basic equations of relativistic cosmology for any value 
of $k$.  We therefore do not agree with the viewpoint that $k=0$ is necessarily
the most natural interpretation of current observational data.  If future
experiments produce a much smaller limit on the flatness parameter
$\varepsilon$ (say, $10^{-5}$), then that might be a more convincing
indication that the most natural value for $k$ is zero.

Chernin has made similar observations, with which we concur \cite{Che03},
although our interpretation is somewhat different.  Chernin emphasizes that
the most notable feature of the Universe is not its near-flatness, nor the
densities of its various components, but the closeness of the Friedmann
constants $C$ and $\Rl$ to each other.  He has subsequently argued that this 
apparent coincidence may in turn be understood most naturally in a closed
universe whose radius of curvature is of the same order of magnitude as
its present size \cite{Che04}.

\section{Summary and Conclusion}

We have presented a new quantitative definition of a nearly-flat Universe
in terms of a flatness parameter $\varepsilon$, and shown that it follows
uniquely from three independent lines of argument based on density,
geometry and dynamical behavior.  It is clear from our derivations that
there is no way, even in principle, to distinguish between a precisely
flat Universe and one whose flatness parameter is sufficiently small.
Measurements of total density from CMB data currently set an upper limit
on $\varepsilon$ of order $10^{-2}$, which is small but not extremely small
or unnatural.  By contrast, the variable quantity $\Ot-1$ is {\em not\/}
a useful flatness criterion, because $\Ot$ is necessarily driven toward
unity at both early and late times (for any value of $k$) by the equations
of relativistic cosmology and the definition of the critical density.

\section*{Acknowledgment}

This work was supported in part by NASA grant 8-39225 to Gravity Probe~B.
We thank our fellow members of the Gravity Probe B Theory Group for many
productive discussions, and V.~Chernin for his useful critical comments.

\end{document}